# The Relations between Agent Performances and Their Intellective Abilities in Mix-games

Chengling Gou

Physics Department, Beijing University of Aeronautics and Astronautics
37 Xueyuan Road, Haidian District, Beijing, China, 100083

Physics Department, University of Oxford
Clarendon Laboratory, Parks Road, Oxford, OX1 3PU, UK
gouchengling@buaa.edu.cn , gouchengling@hotmail.com

Abstract: This paper studies the relations between agent performances and their intellective abilities in mix-games in which there are two groups of agents: one group plays a minority game, and the other plays a majority game. These two groups have different historical memories and different time horizons. It is found that these relations are greatly influenced by the configurations of historical memories of the two groups.

1. Introduction:

Challet and Zhang's minority game (MG) model, together with the original bar model of Arthur [1], attracts a lot of following studies. Given MG's richness and yet underlying simplicity, MG has also received much attention as a financial market model [2~9]. C. Gou proposed an extended model of MG that is referred to as a mix-game in which there are two groups of agents: Group 1 plays a majority game and Group 2 plays a minority game [10, 11].

Among a lot of articles that studied MG or its extensions, only a few papers investigated the relations between agent performances and their intellective abilities. Challet & Zhang [12] and Johnson et al. [13] studied mixed-population MG in which agents have either smaller historical memories or bigger historical memories, and they found that agents with bigger "brain", that is to say, agents have greater historical memories, have advantage to profit from agents with smaller "brain". Is this argument applicable to mix-games? In mix-games, the historical memories of the two groups can be different. C. Gou found that in mix-games if these two groups have the same historical memories, their strategies are anticorrelated; if the historical memory of Group 1 is 6 and greater than that of Group 2, their strategies are uncorrelated; if historical memory of Group 2 is 6 and greater than that of Group 1, their strategies are partly anticorrelated [10]. These results imply that the interplays between these two groups greatly depend on the configurations of their historical memories; i.e. depends on their relative intellective abilities. Therefore, we need to consider agent performances under different configurations of agent historical memories when we study the relations between agent performances and their intellective abilities in mix-games. Through simulations, we find that the relations between agent performances and their intellective abilities in mix-games are greatly influenced by the configurations of historical memories of the two groups: if these two groups have the same historical memories, agent performances of the two groups behave similarly when their historical memories decrease from 6 to 1; if the historical memory of Group 2 is 6, the performances of both groups improve while agents in Group 1 decrease their historical memories from 6 to 1; if the historical memories of Group 1 is 6 and greater than that of Group 2, the performances of these two groups do not influence each other. That is to say, the relations between agent performances and their intellective abilities in mix-games are more complicated than that in mixed-population of MG. This paper is organized as following: in section 2, we describe mix-game models and the simulation conditions; in section 3, the simulation results and discussion are presented; in section 4, the conclusion is reached.

2. Mix-game models and simulation conditions

MG comprises an odd number of agents N choosing repeatedly between the options of buying (1) and selling (0) a quantity of a risky asset. Agents continually try to make minority decisions; i.e. buy assets when the majority of other agents are selling and sell when the majority of other agents are buying. A mix-game model is an extension of MG so its structure is similar to MG. In a mix-game, there are two groups of agents; Group 1 plays a majority game, and Group 2 plays a minority game. N (odd number) is the total number of agents and N1 is the number of agents in Group 1. The system resource is r = 0.5*N. All agents compete in the system for the limited resource. T1 and T2 are the time horizons of the two groups, and m1 and m2 denote the historical memories of the two groups, respectively.

The global information only available to agents is a common bit-string "memory" of m1 or m2 most recent competition outcomes (1 or 0). A strategy consists of a response, i.e., 0 (sell) or 1 (buy), to each possible bit string; hence there are $2^{2^{m1}}$ or $2^{2^{m2}}$ possible strategies for Group 1 or Group 2, respectively, which form full strategy spaces (FSS). At the beginning of a game, each agent is assigned *s* strategies and keeps them unchanged during the game. After each turn, agents assign one virtual point to a strategy that would have predicted the correct outcome. For agents in Group 1, they will reward their strategies one point if they are in the majority side; for agents in Group 2, they will reward their strategies one point if they are in the minority side. Agents collect the virtual points for their strategies over the time horizon T1 or T2, and they use their strategy that has the highest virtual point in each turn. If there are two strategies that have the highest virtual point, agents use coin toss to decide which strategy to be used. This payoff scheme is reasonable and feasible because agents just calculate the virtual points for their strategies as if they used these strategies. The virtual points do not mean the money agents win, but just represent the satisficatory level agents have for their strategies. Reference [14] studies the simulation of financial markets by using mix-game models and specifies the spectra of parameters of mix-game models that fit financial markets according to the dynamic behaviors of mix-game models under a wide range of parameters. The conclusion of reference [14] is that a mix-game model can be a potentially good model to simulate a real financial market.

In simulations, the initial strategy distributions of agents are randomly uniformly distributed in FSS and are kept unchanged during the game. Simulation time-steps are 6000. Number of strategies per agent is 2.

3. Simulation results and discussions

Reference [10] pointed out that N and N1/N influence the dynamic behaviors of mix-games in addition to the configurations of historical memories. Therefore, we did simulations with different N and N1/N under the condition of different configurations of historical memories. Since mix-games display much interesting characteristics within the range of N1/N<0.5 [10, 14], we focused on studying the influence of N1/N within this range.

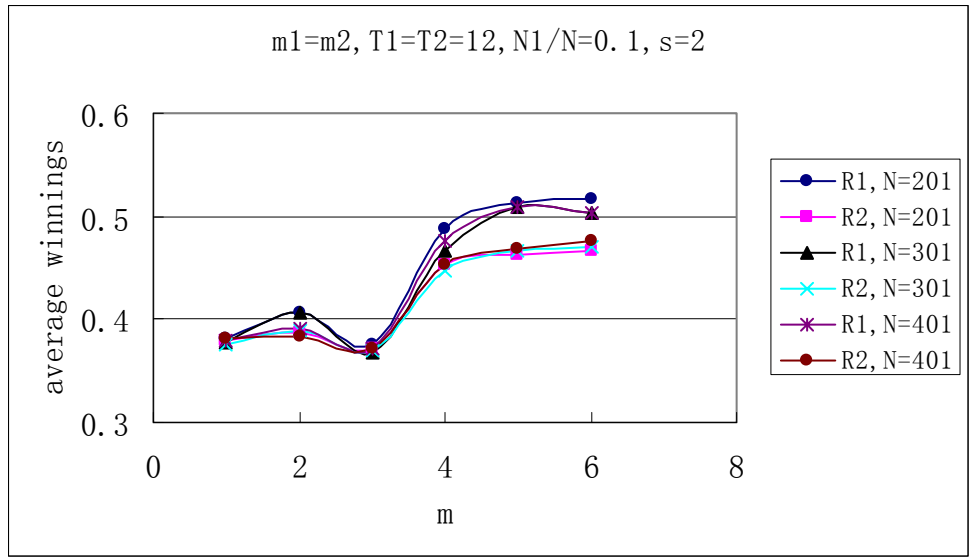

Fig.1a

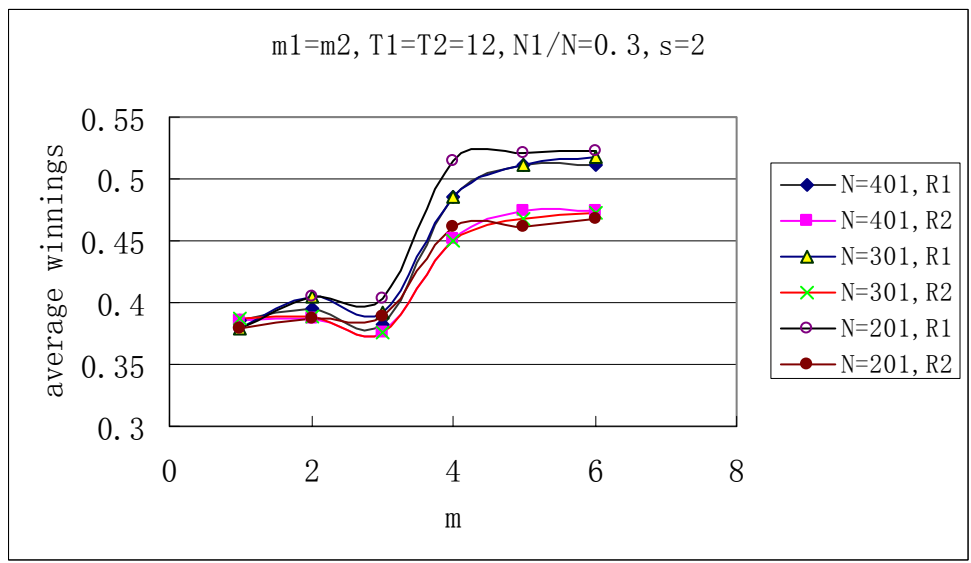

Fig.1b

Fig. 1 the relations between average winnings per agent per turn in mix-games and historical memories of the two groups (m1 and m2) with different N and N1/N under the condition of m1=m2, T1=T2=12, where R1 and R2 represent the average winnings per agent per turn in mix-games for Group 1 and Group 2, respectively.

Figure 1 shows the simulation results about average winnings per agent per turn in mix-games vs. historical memories of the two groups with different N and N1/N under the condition of m1=m2, T1=T2=12, where Fig.1a is for N1/N = 0.1 and Fig.1b is for N1/N = 0.3 respectively. In Fig.1a, the average winnings per agent per turn of the two groups (R1 and R2) behave similarly and have lowest point at m1=m2=3, and they increase greatly at m1=m2=4, and R1 and R2 behave stable if m1=m2>4. Figure 1a also shows that N does almost not influence the relations between average winnings (R1 and R2) and historical memories (m1 and m2). Comparing Fig.1b with Fig.1a, we can say that the relations between average winnings and the historical memories are similar irrespective to the different N1/N. Figure 1 show that if historical memories (m1 and m2) are smaller than 4, the average winnings of these two groups (R1 and R2) are much smaller than 0.5 that is the average winnings for coin-toss.

We show that strategies of Group 1 are fully anticorrelated with that of Group 2 if m1=m2 so that the dynamics of mix-games with m1=m2 is equivalent to that of pure MG with the same parameters but with "the effective number of agents" [10]. In this case, agents in Group 1 match agents in Group 2 to form "anticorrelated pairs". In each "anticorrelated pair", the two agents take the opposite decisions and their payoffs are also opposite so that they get one winning point or not at the same time. This is why R1 and R2 behave similarly.

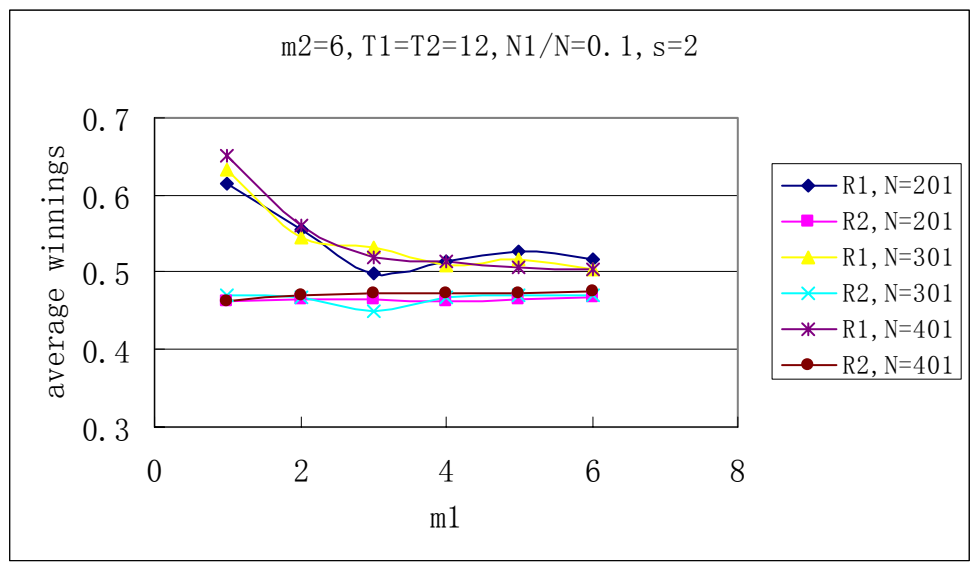

Fig.2a

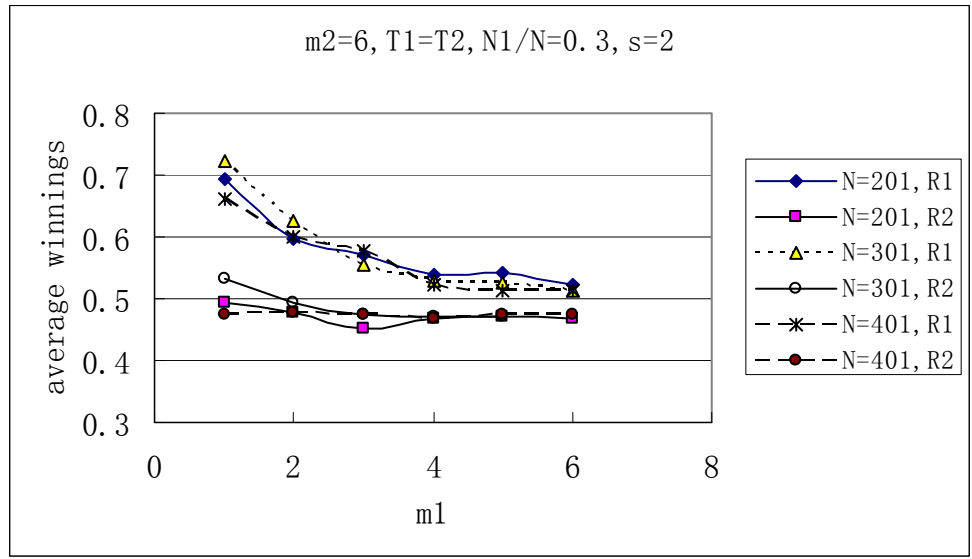

Fig.2b

Fig. 2 the relations between average winnings per agent per turn in mix-games and historical memories of Group 1 (m1) with different N and N1/N under the condition of m2=6, T1=T2=12. R1 and R2 have the same meanings as defined in Fig. 1.

Figure 2 show the average winnings per agent per turn in mix-games vs. historical memories of Group 1 (m1) with different N and N1/N under the condition of m2=6, T1=T2=12, where Fig.2a is for N1/N=0.1 and Fig.2b is for N1/N=0.3, respectively. From Fig. 2b, one can find that the average winnings per agent per turn of both groups (R1 and R2) increase when m1 decreases from 6 to 1 and R1 increases much more quickly than R2. And the average winning of Group 1 (R1) is greater than that of Group 2 (R2). Comparing Fig. 2a with Fig. 2b, we can notice that in Fig.2a R2 behaves relatively stable when m1 decreases from 6 to 1 and R1 increases obviously when m1 is smaller than 3, but the overall relations between average winnings (R1 and R2) and historical memories of Group 1 (m1) are similar in these two cases. That is to say, N1/N just slightly influences these relations, but does not change these relations qualitatively. From Fig. 2, we also can conclude that N does not influence the relations between average winnings (R1 and R2) and historical memories of Group 1 (m1).

Comparing Fig. 2 with Fig. 1, we can notice that R1 and R2 in Fig. 2 behave quite differently from those in Fig. 1 and agents in both groups greatly increase their average winnings under the condition of m1<m2=6. This implies that agents with smaller memories in Group 1 can not only improve their own performance but also benefit for agents in Group 2. This result means that the cooperation between these two groups emerges in this case. This result suggests that the overall performance of the system can be improved under the condition of m1<m2=6.

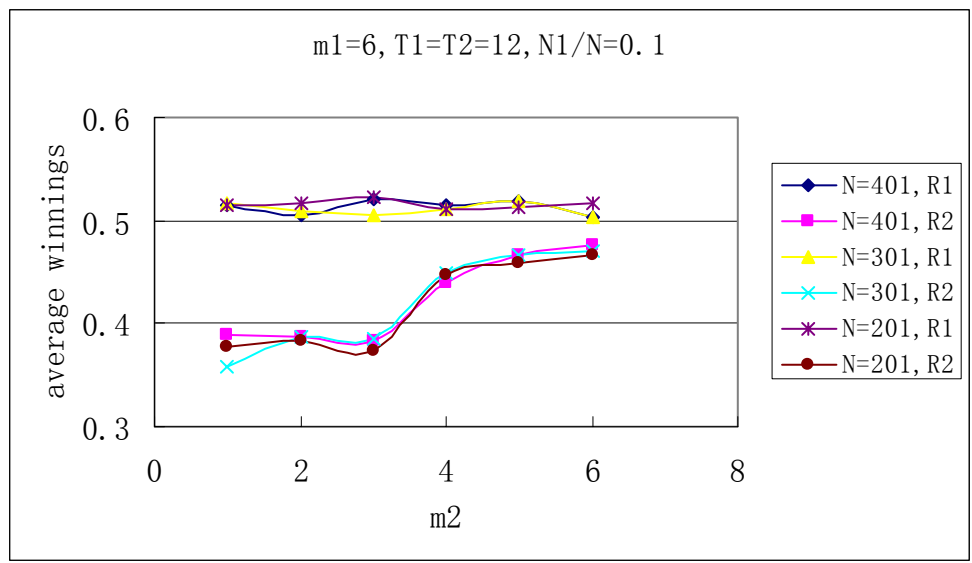

Fig.3a

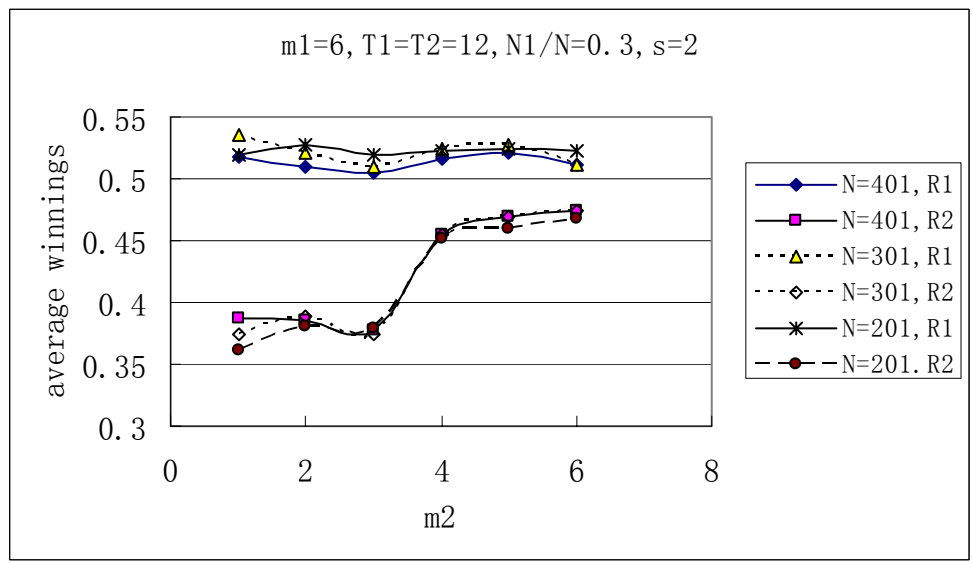

Fig.3b

Fig. 3 the relations between average winnings per agent per turn in mix-game and historical memories of Group 2 (m2) with different N and N1/N under the condition of m1=6, T1=T2=12. R1, R2 and R have the same meanings as defined in Fig. 1.

Figure 3 presents the average winnings per agent per turn in mix-games vs. different memories of Group 2 (m2) with different N and N1/N under the condition of m1=6, T1=T2=12. From Fig. 3a, one can find that the average winning of Group 1 (R1) behaves relatively stable, while the average winning of Group 2 (R2) decreases when m2 decreases from 6 to 1. The average winning of Group 1 (R1) is greater than that of Group 2 (R2). N does not qualitatively influence the relations between average winnings (R1 and R2) and historical memories of Group 2 (m2). Comparing Fig. 3b with Fig. 3a, we can find the overall relations between average winnings (R1 and R2) and historical memories of Group 2 (m2) are similar in these two cases. That is to say, N1/N does not change the relations qualitatively.

Comparing Fig. 3 with Fig. 1, we can see that the relations between average winnings of Group 1 and historical memories of Group 2 (m2) in Fig. 3 are similar to that in Fig. 1. Comparing Fig. 3 with Fig. 2, we can find that the interplays between Group 1 and Group 2 are quit different in these two cases. In Fig. 2, cooperation between the two groups emerged, but in Fig.3, agents in Group 2 can not take advantage of agents in Group 1 and agents in Group 1 also can not take advantage of agents in Group 2. This means these two groups behave independently. This result is in accord with the finding of reference [10].

There is a question that if the above findings depend on parameters of T1 and T2. Through simulations, we find that the above findings don't depend on the parameters of T1 and T2. Figure 4 shows an example of such simulations. Comparing Fig. 4 with Fig. 2b, we can find that the relations between average winnings per agent per turn and the history memory of Group 1 are similar although the time horizons (T1 and T2) are different. This point also holds for parameter configurations of m1=m2 and m1=6>m2.

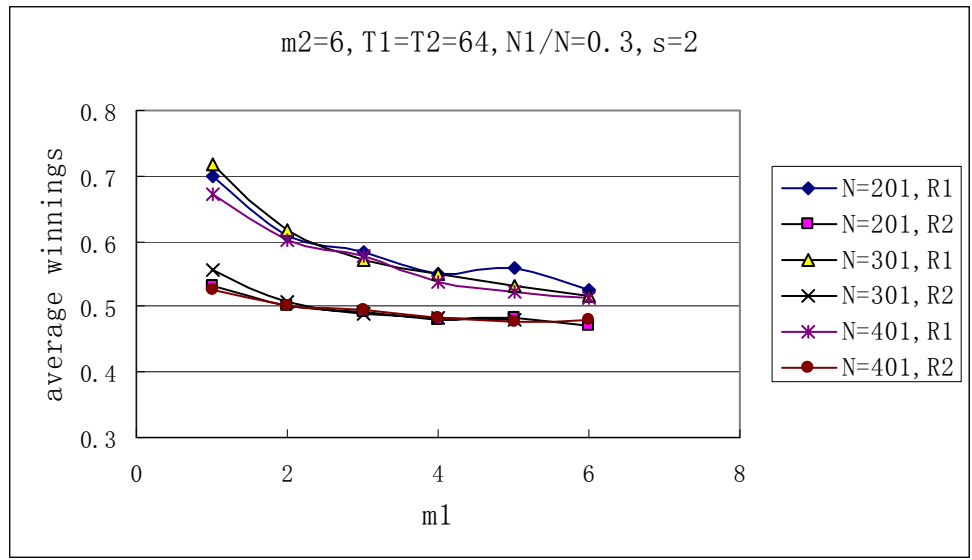

Fig. 4 the relations between average winnings per agent per turn in mix-game and historical memories of Group 1 (m1) under the condition of m2=6, T1=T2=64, N=201, 301, 401 and N1/N=0.3.

4. Conclusion

The relations between agent performances and their intellective abilities in mix-games are greatly influenced by the configurations of historical memories of the two groups. Performance of any group in mix-games not only depends on its own intellective ability, but also greatly depends on that of the other group. These findings do not qualitatively depend on the total number of agents, the time horizons of these two groups and the fraction of agents in Group 1if N1/N<0.5.

Acknowledgements

This research is supported by China Scholarship Council. Thanks Professor Neil F. Johnson for helpful discussing. Thanks David Smith for providing the original program code.